\documentclass[fleqn,usenatbib]{mnras}
\usepackage[T1]{fontenc}
\usepackage{ae,aecompl}
\usepackage{graphicx}   
\usepackage{amsmath}    
\usepackage{amssymb}    
\usepackage{float}

\usepackage[dvipsnames]{xcolor}

\title[Core shift break]%
{On the possible core shift break in relativistic jets}
\author[Nokhrina]{\parbox{\textwidth}{
E.~E.~Nokhrina$^{1}$\thanks{E-mail: nokhrinaelena@gmail.com}
}
\vspace{0.4cm}\\
\parbox{\textwidth}{
$^1$Lebedev Physical Institute, Leninsky prosp.~53, Moscow, 119991, Russia
}
}

\begin{document}

\date{Accepted 2024 November 5. Received 2024 November 5; in original form 2024 June 23}

\pagerange{\pageref{firstpage}--\pageref{lastpage}} \pubyear{}

\maketitle

\label{firstpage}

\begin{abstract}
Measurement of a jet geometry transition region is an important instrument of assessing the jet ambient medium properties, plasma bulk motion acceleration, parameters of a black hole and location of a jet launching radius. In this work we explore the possibility of a presence of a core shift break, associated with the geometry and jet physical properties transition. We obtain the relations on the core shift offset jump due to a change in a core shift exponent. The condition of a proper frame magnetic field continuity and the core shift break can be used as an instrument to refine the magnetic field estimates upstream the break. This method is applied to the jet in NGC~315. We speculate that the localised in a flow plasma heating either by reconnection or due to particles acceleration at the shock will also lead to a core shift break, but of a different type, than the one observed in NGC~315. We propose to use the multi-frequency core shift measurements to increase the number of sources with a detected jet shape break and to boost the accuracy of assessing the properties of a jet geometry transition region.
\end{abstract}


\section{Introduction}
\label{s:intro}

Since the first discovery of a jet shape break by \citet{Asada12} a lot of work has been done in order to understand the physics of this effect, and to apply it to assessing the properties of relativistic jets in active galactic nuclei (AGN).  
For many years jets have been thought of as a conical outflows, with the jet apparent opening angles being measured for a great number of sources \citep[see e.g.][and references within]{Pushkarev_09}. Later the multi-epoch stacking technique has been applied to ensure the full jet width is accounted for \citep{MOJAVE_XIV}. Both one-epoch and multi-epoch observations revealed for the closest sources the phenomenon of a jet shape break --- a transition of an observed jet boundary shape from a quasi-parabolic to a quasi-conical \citep{Asada12, tseng16, Hada18, r:Akiyama18, r:Nakahara18, r:Nakahara19, r:Nakahara20, Kov-20, Park_2021, Boccardi_2021, Okino22}.

Jet shape on different scales may be a probe for both jet pressure at its boundary and an ambient medium pressure, and a key to understanding the FRI--FRII dichotomy. \citet{Asada12} discuss different models for a jet pressure, possible distributions of ambient pressure and their connection with knots HST-1 and A in M~87. \citet{Kov-20} also note the possible connection of a jet shape break with a stationary bright feature that can be a standing shock, appearing due to a change in an ambient medium pressure profile. \citet{GL17}, using the results by \citet{Lyu09}, showed, that both the observed boundary shape in M~87 jet and the position of HST-1 can be explained by introducing the abrupt change in an ambient medium pressure, although with some caveats, that they discuss. The change in an ambient medium profile is expected to occur around a Bondi radius or a radius of a sphere of gravitational influence (SGI) \citep{Asada12, BMR-19}. At the same time, most of the sources have a break at the radius $\sim 10^5-10^6\;r_\mathrm{g}$ \citep{Asada12, Hada18, Kov-20}, although in the case of 1H~0323$+$342 the mass estimate is ambiguous \citep{Hada18, NKP20_r2}. On the other hand, two sources with the detected jet shape break deviate from this trend. Independent evaluation of a geometry transition in NGC~315 jet by \citet{Park21, Boccardi_2021} located the break one to two orders of magnitude in length inside the Bondi sphere. The break position in 3C~273 is detected an order of magnitude outside the SGI \citep{Okino22}. These sources require disk winds, torus or hot cocoon surrounding a jet \citep{Bromberg11} to account for a change in an ambient medium pressure profile that can lead to a jet shape break. 

The connection between a transition from a quasi-parabolic to a quasi-conical jet boundary shape and a plasma acceleration profile has been proposed by \citet{Lyu09, Hada18, Kov-20}. Analytical solution by \citet{Lyu09} predicted such a transition when the flow becomes particle-dominated. In fact, \citet{Kom07} using numerical simulations noted the smooth  evolution of a pressure profile at the jet boundary. \citet{BCKN-17, Kov-20} obtained the change in a jet pressure and the corresponding change in a jet boundary shape as the flow accelerates up to half of maximum Lorentz factor. As it was shown by \citet{Tom_tak_2003, Beskin06, Kom07, Lyu09, TMN09}, a jet accelerates with the bulk Lorentz factor of plasma motion $\Gamma$ growing almost linearly with the jet width $d$ as
\begin{equation}
\Gamma\approx\frac{d/2}{R_\mathrm{L}}.
\label{acc_ideal}
\end{equation}
Here the light cylinder radius $R_\mathrm{L}=c/\Omega_\mathrm{F}$ is defined by the angular velocity $\Omega_\mathrm{F}$ constant at least in the inner part of a jet \citep{Lyu09}. After reaching the value $\Gamma_\mathrm{max}/2$, Lorentz factor growth saturates and becomes logarithmically slow \citep{Beskin06}. $\Gamma_\mathrm{max}$ corresponds to the maximum bulk motion Lorentz factor if all the electromagnetic energy is transformed into plasma kinetic energy. Depending on the normalisation, it is proportional to the jet initial magnetisation $\sigma_\mathrm{M}$ with a factor of a few \citep{Beskin06, NKP20_r2}.
The pressure on a jet boundary is different for a magnetically dominated and plasma dominated regimes \citep{Vlahakis04, N24}. Thus, the single ambient medium pressure power law dependence on a distance along a jet, consistent with Bondi accretion, can account for the break in a jet shape. The connection between this break and an acceleration saturation has been explored by \citet{Hada18, Nakamura+18, Kov-20, Ricci22}. This connection allows estimating such properties of a jet as a light cylinder radius and ambient medium pressure amplitude \citep{NKP20_r2}. 

The transformation of a Poynting flux into plasma bulk motion kinetic energy flux is not the only structural change of a jet that influences its pressure at the boundary. The appearance of a dense core with almost uniform plasma density and a poloidal magnetic field has been predicted for high enough ambient pressure, expected on the sub-parsec scales \citep{Kom09, Beskin09}. The change in a jet pressure connected with this transformation leads to appearance of an inner break and has been confirmed in M~87 jet on the scales of the order of a few milliarcseconds from the jet base \citep{Beskin24}.

Extrapolating the boundary field line down to the gravitational radius scale may provide an answer to a jet origin: \citet{Asada12, Nakamura+18, Nokhrina2022} argue that the jet in M~87 is launched from the regions nearby the central black hole, suggesting that the main jet power is a Blandford--Znajek process \citep{BZ-77}. It is worth noting that to do this, it is important to fit not only the power of jet shape on small scales, but also the position of an apex of fitted quasi-parabola \citep{Hada11_M87, Kov-20}. 
 
There are limitations on detecting the phenomenon of a jet shape break directly due to the finite resolution of existing instruments. Indeed, jet shape break has been observed directly only for a dozen nearby sources (typical red shift of the order of a few tenth or less), and those observed at relatively large viewing angles (for example, for BL Lac the viewing angle $\theta=7^{\circ}.6$). The last fact is connected with the resolution along a jet. Due to the same issue, the breaks has been observed up to now, probably, mainly in jets with large enough light cylinder radius. Indeed, if the break is explained by a conversion of a half of a Poynting flux into a plasma bulk motion kinetic energy flux, then, for the observed maximum Lorentz factors $\lesssim 50$ \citep{MOJAVE_XVII}, the jet width at the break is of the order of $100-200 R_\mathrm{L}$. The cited above works provide the observed jet width at the break to be of the order of a few tenths of a parsec, which leads to rather large light cylinder radii $\sim 10^{-3}$~pc \citep{Nokhrina19, NKP20_r2}. The corresponding estimate of a light cylinder radius for the jet in M~87 was confirmed independently by modelling analytically the kinematics by \citet{Kino22}. Up to now only the jet in NGC~315 has been resolved so finely, that the implied dimensionless black hole spin is close to unity, which means that the light cylinder radius constitutes a few gravitational radii \citep{Ricci22}. All these circumstances pose limitations on studying the jet shape break phenomenon directly on a larger sample of sources.

There can be an indirect way to detect the effect in more distant jets or jets, observed at smaller observational angles. It was shown that the plasma bulk Lorentz factor, estimated by the delays in flares from radio cores at 8 and 15~GHz \citep{KGAA2011, kutkin19} and the position of these cores, obtained using core shift method, are consistent with a parabolic jet shape for the sources with red shifts up to $\sim 2$ \citep{Nokhrina2022}. However, this method requires multi-epoch observations, high variability of a source and do not provide directly the position of a break. 

We discuss here the possibility of detecting a jet shape break allowed that it is also seen as a break in an apparent (projected) core position $r_\mathrm{core}$ as a function of observational frequency $\nu$. First of all, there are now many indications, that the core shift exponent $k_\mathrm{r}$ in the dependence
\begin{equation}
r_\mathrm{core}=\Omega_{r\nu}\nu^{-1/k_\mathrm{r}},
\label{r_core}
\end{equation}
can deviate from the unity, expected in conical non-accelerating jets \citep[see e.g.,][and the references within]{Nokhrina2022}. Here $\Omega_{r\nu}$ is a core shift offset, which depends on the physical properties affecting the synchrotron opacity at the core.
As the dependence of an apparent core shift on the observational frequency is different for different flow regimes (magnetically/plasma dominated) and for different observed boundary shapes \citep{Nokhrina2022, Ricci22, N24}, one may expect to detect its break. We explore this possibility in this paper. We use the results from the papers \citet{Ricci22} and \citet{N24} and refer to them as R22 and N24 for brevity everywhere below.

The paper is organised as follows. In section~\ref{s:sec1} we develop the theoretical basis and discuss the break in a core shift offset. Assuming the magnetic field continuity, we obtain the formulas that can be applied to the particular sources in section~\ref{s:magfield}. We show the results for NGC~315 in section~\ref{s:ngc315}. Finally, we discuss the results and possible applications of a method in~\ref{s:Dis} and summarise our work in section~\ref{s:Conc}.

\section{Break in a core shift}
\label{s:sec1}

In general, a jet can be divided in tree regions depending on its local shape: quasi-parabolic, quasi-conical and more or less smooth transition between them. In this work we regard for simplicity two asymptotic power laws (parabolic and conical). We extrapolate them into a transition region, so they intersect at the distance $r_\mathrm{br}$. Thus, we model the jet boundary shape by the function
\begin{equation}
d=d_\mathrm{br}\left(\frac{r}{r_\mathrm{br}}\right)^{k},\quad k=\left\{
\begin{array}{l}
k_1,\quad r\le r_\mathrm{br},\\
k_2,\quad r>r_\mathrm{br},
\end{array}
\right.
\label{shape}
\end{equation}
where $d$ is the jet width at the distance $r$ along a jet, and $k$ is an exponent that defines the particular boundary shape. Upstream the break an outflow sustains a quasi-parabolic shape with $k_1\approx 0.5$, while downstream an exponent $k_2\approx 1$ corresponds to a quasi-conical flow. Here and below $r$ is a distance from the jet base. For the radio core given by Equation~(\ref{r_core}) $r=0$ corresponds to the infinite observational frequency. 

The bulk plasma motion Lorentz factor $\Gamma$ affects the magnitudes of an emitting plasma number density and a magnetic field in a plasma proper frame. Magnetohydrodynamical modelling provides three distinct phases of its growth: the effective acceleration given by Equation~(\ref{acc_ideal}), the transitional domain, where the acceleration slows down, and the saturation domain, where the bulk Lorentz factor grows logarithmically slow with a distance \citep{Tom_tak_2003, Beskin06, Kom09, Lyu09, TMN09}. We connect the jet boundary shape with the Lorentz factor behaviour \citep{BCKN-17, Nokhrina19, Kov-20, Ricci22}. Assuming for simplicity two asymptotic domains, we set the following dependence of a maximum bulk Lorentz factor on a local jet width $d$ for a given jet cross-cut:
\begin{equation}
\Gamma=\left\{
\begin{array}{l}
\displaystyle\rho \frac{d}{2R_\mathrm{L}},\quad d\le d_\mathrm{br},\\ \ \\
\displaystyle\frac{\Gamma_\mathrm{max}}{2},\quad d>d_\mathrm{br}.
\end{array}
\right.
\label{G}
\end{equation}
In the region $r\le r_\mathrm{br}$ ($d\le d_\mathrm{br}$) a jet is dominated by the Poynting flux and effectively accelerates. 
The coefficient $\rho$ defines how strongly the Lorentz factor of a bulk plasma motion deviates from the ideal linear acceleration given by Equation~(\ref{acc_ideal}). It accounts for both a transition between acceleration and saturation domains and a possible presence of a slow sheath. Typical values of $\rho$ from analytical and numerical simulations are $0.20-0.55$ \citep{Kom07, Chatterjee2019, BCKN-17, Kov-20}. 
Further downstream we assume that an outflow is dominated by a plasma bulk motion kinetic energy flux, an acceleration saturates, and a jet assumes a quasi-conical shape, described by the Equation~(\ref{shape}), with $k_2=1$ for a strict cone.
Here we must emphasize that the detected break in a jet shape does not necessarily coincides with the boundary between the Poynting -- kinetic energy flux dominance (see e.g. Figure~7 in \citet{Kov-20}). However, we assume that both are located in the transition region. As we regard two asymptotic branches, we set $r_\mathrm{br}$ and $d_\mathrm{br}$ to represent both a break in a jet shape and a flow transition from effective to saturated acceleration. We denote all the values upstream the break in a core shift with an index ``1'', downstream --- with an index ``2'', and in a plasma proper frame --- with an asterisk.

We assume the power law dependencies of physical quantities that define synchrotron opacity on a jet width $d$ or, equally, on a distance along a jet $r$. A dominant magnetic field $B_*$ and an emitting plasma number density amplitude $K_\mathrm{e*}$ in a plasma proper frame are defined by the expressions 
\begin{equation}
B_*=B_{*\mathrm{br}}\left(\frac{r_\mathrm{br}}{r}\right)^{k_\mathrm{b}},\quad k_\mathrm{b}=\left\{
\begin{array}{l}
k_\mathrm{b1},\quad r\le r_\mathrm{br},\\
k_\mathrm{b2},\quad r>r_\mathrm{br},
\end{array}
\right.
\label{defB}
\end{equation}
and 
\begin{equation}
K_\mathrm{e*}=K_\mathrm{e*\mathrm{br}}\left(\frac{r_\mathrm{br}}{r}\right)^{k_\mathrm{n}},\quad k_\mathrm{n}=\left\{
\begin{array}{l}
k_\mathrm{n1},\quad r\le r_\mathrm{br},\\
k_\mathrm{n2},\quad r>r_\mathrm{br}.
\end{array}
\right.
\label{defn}
\end{equation} 
Here the exponents $k_\mathrm{n}$ and $k_\mathrm{b}$ may in general change at the $r_\mathrm{br}$. We fix the values of a magnetic field and an emitting plasma number density amplitude at the distance $r_\mathrm{br}$ as $B_{*\mathrm{br}}$ and $K_\mathrm{e*\mathrm{br}}$ correspondingly. The emitting plasma number density has an energy distribution $\mathrm{d}n_*=K_\mathrm{e*}\gamma^{-1-2\alpha}\mathrm{d}\gamma$, $\gamma\in[\gamma_\mathrm{min},\;\gamma_\mathrm{max}]$, with an exponent defined by the spectral index $\alpha$ in an optically thin regime. It is related to its amplitude $K_\mathrm{e*}$ by the standard expression $n_*\approx K_\mathrm{e*}/2\alpha\gamma_\mathrm{min}^{2\alpha}$.We set the broken power law dependence of a Doppler factor
\begin{equation}
\delta=D\left(\frac{r}{r_\mathrm{br}}\right)^{\mathrm{k}_{\delta}}
\label{dlt}
\end{equation}
with $k_{\delta 1}\ne k_{\delta 2}$ in general. For the observational angle $\theta<\Gamma_\mathrm{max}$ one can set $\delta(r)\approx 2\Gamma(r)$, so that 
\begin{equation}
\delta=\Gamma_\mathrm{max}\left(\frac{r}{r_\mathrm{br}}\right)^{\mathrm{k}_{\delta}},\quad \mathrm{k}_{\delta}=\left\{
\begin{array}{l}
k,\quad r\le r_\mathrm{br},\\
0,\quad r>r_\mathrm{br}.
\end{array}
\right.
\label{dlt-1}
\end{equation} 
If a jet is observed at the angle larger than $\Gamma_\mathrm{max}$, we detect predominantly the slower parts of an outflow with $\Gamma\lesssim\theta^{-1}$. In this case one can set $\delta=\mathrm{const}$ everywhere along a jet.

Equations~(\ref{shape})--(\ref{dlt-1}) provide the projected position $r_{\mathrm{core}}$ of a radio core --- a surface where an optical depth is equal to unity --- to be a power law function of an observational frequency $\nu$ given by the Equation~(\ref{r_core}) with
\begin{equation}
k_\mathrm{r}=\frac{k_\mathrm{n}+(1.5+\alpha)(k_\mathrm{b}-k_{\delta})-k}{2.5+\alpha}.
\label{kr}
\end{equation}
The Equation~(\ref{r_core}) holds in two separate domains, where $d(r)$, $\Gamma(r)$, $B_*(r)$, $K_\mathrm{e*}(r)$ and $\delta(r)$ can be described by a single power law.

It follows from the plasma number density flux continuity that in the effective acceleration regime (the first line of the Equation~(\ref{G})) $k_\mathrm{n1}=3k$, while in the
saturation regime (the second line of the Equation~(\ref{G})) $k_\mathrm{n2}=2k$ (see details in R22 and N24). In the effective acceleration domain the plasma proper frame magnetic field is either dominated by the poloidal component or the latter is of the same order as the toroidal component \citep{Vlahakis04, Kom07}. In the saturation domain the toroidal component dominates in the plasma proper frame. Thus, we use $k_\mathrm{b1}=2k$ and $k_\mathrm{b2}=k$ correspondingly. In particular, for a conical jet boundary geometry with $k=1$, $k_\mathrm{n2}=2$, $k_\mathrm{b2}=1$ and $k_{\delta}=0$ in accordance with \citet{BlandfordKoenigl1979}.
The described above differences in the exponents affect the values of the core shift exponent $k_\mathrm{r}$ in the Equation~(\ref{kr}). In a plasma-dominated regime $k_\mathrm{r2}=k_2$ (R22).
In a Poynting-dominated domain 
the core shift exponent $k_\mathrm{r1}=2k_1$ if a jet is observed at a viewing angle $\theta\approx\Gamma^{-1}$ (Doppler factor $\delta\approx\mathrm{const}$, see N24).
For a smaller viewing angle
$\theta\ll\Gamma^{-1}$ ($\delta\approx 2\Gamma$) the core-shift exponent is defined by the expression $k_\mathrm{r1}=k_1(3.5+\alpha)/(2.5+\alpha)$. 

Measuring in milliarcseconds the shift $\Delta r_\mathrm{mas}$ between the core positions at two different observational frequencies $\nu_1$ and $\nu_2$, we calculate the core shift offset
\citep{L98}
\begin{equation}
\Omega_{r\nu}=4.85\times 10^{-9}\frac{\Delta r_\mathrm{mas}D_\mathrm{L}}{(1+z)^2}\left(\frac{\nu_1^{1/k_\mathrm{r}}\nu_2^{1/k_\mathrm{r}}}{\nu_2^{1/k_\mathrm{r}}-\nu_1^{1/k_\mathrm{r}}}\right)\;\mathrm{pc\, GHz^{1/k_\mathrm{r}}}.
\label{Omega}
\end{equation}
Here a source is located at the luminosity distance $D_\mathrm{L}$ and corresponding cosmological red shift $z$.
On the other hand, the modelling shows that a core shift offset depends on a jet geometry and its physical properties \citep{L98, hirotani2005, N24}: 
\begin{equation}
\left(\frac{\Omega_{r\nu}}{\sin\theta}\right)^{p}=\frac{C(\alpha)d_\mathrm{br}}{\sin\theta}\left(\frac{D}{1+z}\right)^{1.5+\alpha} r_{\mathrm{br}}^{p}
K_\mathrm{e*\mathrm{br}}B_{*\mathrm{br}}^{1.5+\alpha}.
\label{Ojump}
\end{equation}
Here 
\begin{equation}
p=k_\mathrm{n}+(1.5+\alpha)(k_\mathrm{b}-k_{\delta})-k,
\label{p_power}
\end{equation}
and we set the reference distance for determining $B_*$ and $K_\mathrm{e*}$ as $r_\mathrm{br}$ instead of 1~pc. Jet width $d_\mathrm{br}$ and a break position along a jet $r_\mathrm{br}$ here and below are in parsecs. The term $C(\alpha)$ is a dimensional constant, which depends on the emitting plasma energy distribution, and $C(0.5)=580\; \mathrm{cm}^4\;(\mathrm{s\; g\; pc})^{-1}$. 

\begin{figure}
    \centering
    \includegraphics[width=0.9\linewidth]{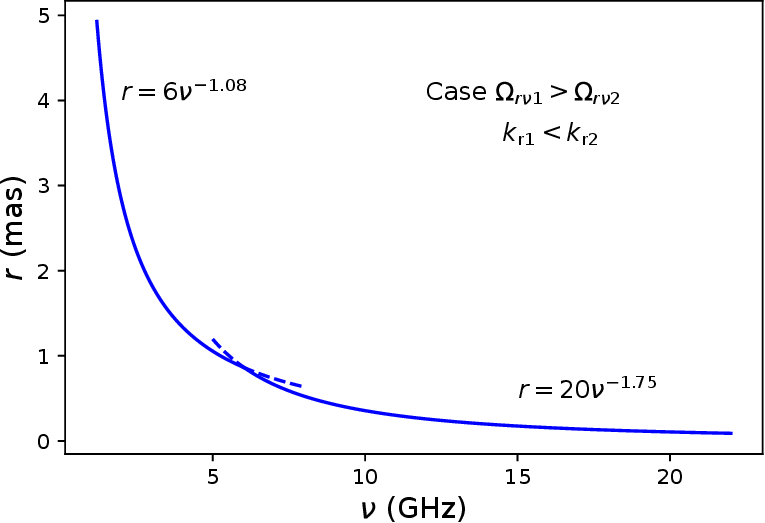} \vspace{0.1cm}
    \includegraphics[width=0.9\linewidth]{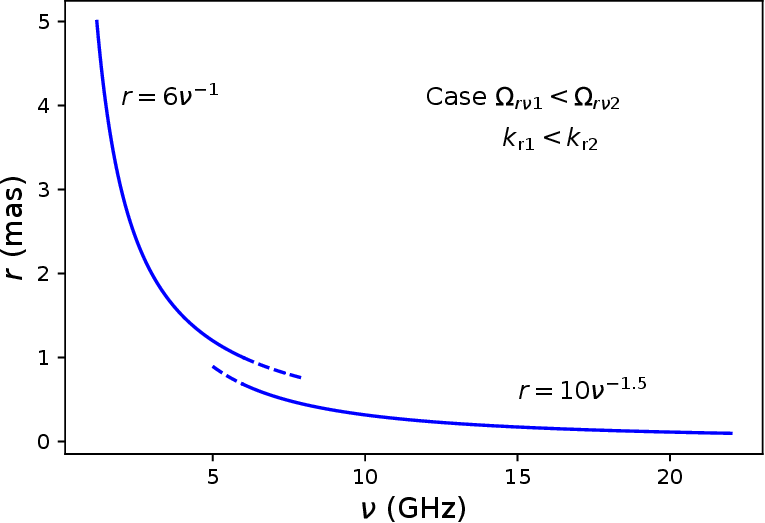}
    \vspace{0.1cm}
    \includegraphics[width=0.9\linewidth]{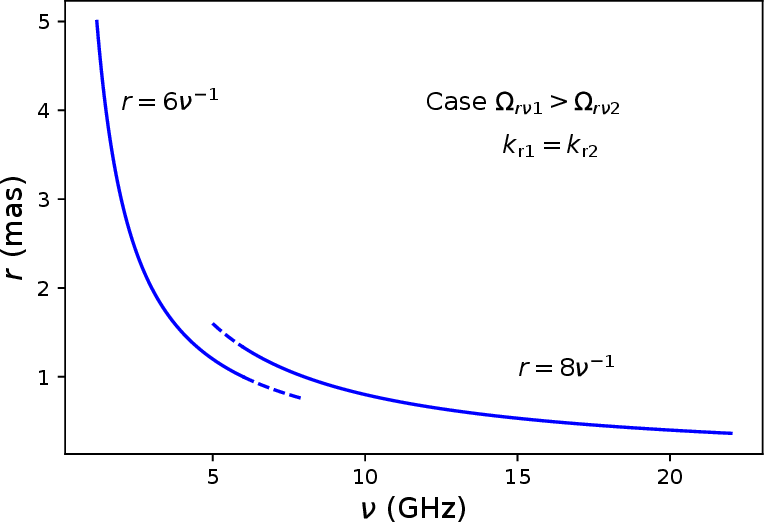} 
    \caption{Cartoons of the possible scenarios of a core shift break. Upper panel: case of $\Omega_{r\nu1}>\Omega_{r\nu2}$ and $k_\mathrm{r1}<k_\mathrm{r2}$ as observed in NGC~315.
    Medium panel: case of $\Omega_{r\nu1}=\Omega_{r\nu2}$ or $\Omega_{r\nu1}<\Omega_{r\nu2}$ and $k_\mathrm{r1}<k_\mathrm{r2}$. Lower panel: case of $\Omega_{r\nu1}>\Omega_{r\nu2}$ and $k_\mathrm{r1}=k_\mathrm{r2}$, possible for a jump in $K_{e*}$ and $B_*$. Extrapolation of each branch onto adjacent region is shown in dashed lines. Here $r$ is measured from the jet base, assuming its coincidence with the radio core position at infinite frequency.}
    \label{f:cartoons}
\end{figure}

The exponents $k_\mathrm{n}$, $k_{\mathrm{b}}$, $k_{\delta}$ and $k$ assume different values in different domains along a jet. We emphasize that the power law relation~(\ref{r_core}) holds only in the part of a jet, which is described by a single power law for each physical value $B_*(r)$, $K_\mathrm{e*}(r)$, $\delta(r)$ and $d(r)$.
This may not be a case in the transition region, where the core shift position as a function of an observational frequency changes from the one asymptotic power law to the other.
We regard separately quasi-parabolic and quasi-conical domains, where the core shift exponents may differ from each other: $k_\mathrm{r1}\ne k_\mathrm{r2}$ (N24). The core shift offsets $\Omega_{r\nu 1,\,2}$ can be introduced only within these regions correspondingly. In general, $\Omega_{r\nu 1}\ne \Omega_{r\nu 2}$, and it is to this difference in two physically and geometrically different jet domains we refer as a `break in a core shift offset'. 
The two branches described by the Equation~(\ref{r_core}) can also be extrapolated onto the region of a smooth transition: see, for example, the upper panel in Figure~\ref{f:cartoons}.

Let us discuss different possibilities of a jet boundary shape break and its impact on the value of the core shift offset. Equation~(\ref{Ojump}) shows that if there is no discontinuities in a jet properties (velocity, emitting plasma number density and magnetic filed), than a core shift offset may assume different values in two distinct region only if 
$p_1\ne p_2$. 
From the Equations~(\ref{kr}) and (\ref{p_power}) we obtain that $k_\mathrm{r1}\ne k_\mathrm{r2}$ either. Then, for any acceleration regime, Equation~(\ref{Ojump}) provides the jump in a core shift offset for two asymptotic domains, and its value is given by
\begin{equation}
\begin{array}{rcl}
\displaystyle\frac{\Omega_{r\nu1}}{\Omega_{r\nu2}}&=&\displaystyle\left(\frac{\Omega_{r\nu2}}{r_\mathrm{br}\sin\theta}\right)^{(k_\mathrm{r2}/k_\mathrm{r1}-1)}=\\ \ \\
&=&\displaystyle\left(\frac{\Omega_{r\nu1}}{r_\mathrm{br}\sin\theta}\right)^{(1-k_\mathrm{r1}/k_\mathrm{r2})}.
\end{array}
\label{Jump}
\end{equation}
There can be several different physical cases underlying such a jump. 

{\it Pure jet shape change within one of the regimes of plasma acceleration.} Suppose the jet changes its shape at the distance $r_\mathrm{br}$ due to a change in an ambient medium pressure, i.e. $k_1\ne k_2$. However, the plasma acceleration regime is the same: either effective or saturated acceleration. The powers $k_\mathrm{n}$, $k_\mathrm{b}$, $k_{\delta}$ are different in the regions $r<r_\mathrm{br}$ and $r>r_\mathrm{br}$, as the physical values depend on a jet width $d(r)$.
Consequently, $k_\mathrm{r}$ and $p$ change, leading to two different power laws with different core shift offsets.

{\it Pure change of acceleration regime without a jet shape change.} Although $k_1=k_2$ in this case, the exponents $k_\mathrm{n}$, $k_\mathrm{b}$, $k_{\delta}$ change as the outflow transits from Poynting to kinetic energy flux dominance. The exponents $k_\mathrm{r}$ and $p$ are also different, and $\Omega_{r\nu 1}\ne\Omega_{r\nu 2}$.

{\it Jet changes its shape and plasma acceleration regime simultaneously.} 
In the case of $\theta\ll \Gamma^{-1}$ ($\delta\approx 2\Gamma$, see Equation~(25) in N24) the powers $k_\mathrm{r}$ are, in general, not equal to each other even for ideal situation $k_1=0.5$, $k_2=1$: $k_{r1}=4k_1/3\approx 0.66$ and $k_{r2}=k_2\approx 1$. Thus, the jump in core shift offset is always present in this case. 

Suppose the core shift exponents have approximately equal values. This can take place for a jet observed at large enough viewing angle, so that we can set Doppler factor to be approximately constant. In the ideal situation $k_1=0.5$, $k_2=1$, $k_\mathrm{r1}=2k_1$ and $k_\mathrm{r2}=k_2$. In this case the equality $k_\mathrm{r1}\approx k_\mathrm{r2}$ holds, and the jump in core shift offset is absent: although such physical values as emitting plasma number density and magnetic field have different dependencies on $r$ in two different regions, both branches of a core shift coincide, ensuring the same value of $\Omega_{r\nu 1}=\Omega_{r\nu 2}$ along a jet.

In Figure~\ref{f:cartoons} we present the cartoon of possible asymptotic core shift branches behaviour. In general, we expect $k_\mathrm{r1}\le k_\mathrm{r2}$ for a flow that is better collimated at its base. The upper panel demonstrates the case of NGC~315, in which two asymptotic branches intersects, which holds while $\Omega_{r\nu 1}>\Omega_{r\nu 2}$. We do not exclude the opposite case $\Omega_{r\nu 1}<\Omega_{r\nu 2}$, presented in the medium panel. Here we predict two non-intersecting branches.

{\it Case of a jump in plasma properties.} Suppose now that neither a jet shape nor plasma acceleration regime changes, so the core shift exponent $k_\mathrm{r}$ is the same everywhere. However, there is a jump in emitting plasma number density $K_{e*}$ and magnetic field $B_*$. This can be a consequence of either a plasma heating or a propagating flare. In general, the jump in core shift offset in this case is equal to 
\begin{equation}
\frac{\Omega_{r\nu1}}{\Omega_{r\nu2}}=\left(\frac{K_{e*1}B_{*1}^{1.5+\alpha}}{K_{e*2}B_{*2}^{1.5+\alpha}}\right)^{1/p}.
\label{Flare}
\end{equation}
Example of such a core shift break is presented in Figure~\ref{f:cartoons}, the lower panel. In this case the core shift exponents are the same, but the jump in the offset is due to abruptly changed balance between a magnetic field and an emitting plasma energy densities.

The jump in a core shift offset must be expected in jets. For the continuous flow properties its presence is a consequence of different values of a core shift exponent $k_\mathrm{r}$ in different regions of a jet. If it is caused by the jet shape and / or acceleration regime break, its value is universally described by Equation~(\ref{Jump}). In this case we cannot distinguish whether a jump in a  core shift offset is present because of a change in a plasma acceleration pattern or because of a change in a jet shape due to different properties of a collimating medium. If the jump in a core shift offset is due to a change in emitting plasma number density / magnetic field (see Equation~(\ref{Flare})), than we expect the same value of the exponent $k_\mathrm{r}$. The value of the jump, in general, would not obey the Equation (\ref{Jump}), which allows distinguishing the possible effect of a flaring event on a core shift.   
\section{Matching magnetic field at the break}
\label{s:magfield}

Let us assume that the multi-frequency measurements of a core shift are consistent with a break in the Equation~(\ref{r_core}) at some break frequency $\nu_\mathrm{br}$. In particular, suppose that we measure $k_\mathrm{r1}$ for $\nu>\nu_\mathrm{br}$ and $k_\mathrm{r2}$ for $\nu<\nu_\mathrm{br}$. In this case there is in general a jump in the core shift offset as well. Below we regard the simultaneous change in both the plasma acceleration regime and in jet boundary shape as a cause. 

The corresponding $\Omega_{r\nu1}$ and $\Omega_{r\nu2}$ can be calculated using the observational data in two separate jet parts. 
Correspondingly, the frequencies, between which one can calculate the core shift offset, are divided in two sets: for the cores situated in one asymptotic region, and in the other. Mixing the frequencies may lead to unreliable results for the core shift offset. From each set of observational frequencies and the definition given by the Equation~(\ref{r_core}) we readily obtain the de-projected position of an expected jet shape break in parsecs as
\begin{equation}
r_\mathrm{br}=\frac{\Omega_{r\nu1}}{\nu_\mathrm{br}^{1/k_\mathrm{r1}}\sin\theta}.
\label{r}
\end{equation}
Here $\nu_\mathrm{br}$ is in GHz and $\Omega_{r\nu 1}$ is in $\mathrm{pc\, GHz^{1/k_\mathrm{r1}}}$.
If the apparent opening angle $\varphi_\mathrm{app}=2\varphi/\sin\theta$ is measured at the frequencies $\nu>\nu_\mathrm{br}$ (the expected conical part of a jet), one can use the expression
\begin{equation}
d_\mathrm{br}= \varphi_\mathrm{app}\,\frac{\Omega_{r\nu1}}{\nu_\mathrm{br}^{1/k_\mathrm{r1}}}
\label{d_proxy}
\end{equation}
as a proxy for a jet width at the break. Both estimates given by Equations~(\ref{r}) and (\ref{d_proxy}) may be approximate if the intersection of two branches of a core shift do not coincide precisely with the jet geometry transition distance $r_\mathrm{br}$.

Having two branches of the measured core shift, corresponding to two different jet boundary shapes, one can match the corresponding magnetic fields, calculated in two different regions.

Suppose first, that the break in a core shift dependence on the frequency has been detected together with a break in a jet shape. In this case we have the following set of measured parameters: the position of a break along a jet $r_\mathrm{br}$, the jet width at the break $d_\mathrm{br}$, jet shape exponents $k_1$ in a quasi-parabolic domain and $k_2$ in a quasi-conical domain, $k_\mathrm{r1}$ and $k_\mathrm{r2}$ --- corresponding powers in Equation~(\ref{r_core}), $\Omega_{r\nu1}$
and $\Omega_{r\nu2}$ --- core shift offsets in two domains. Let us obtain the conditions, under which the estimates for the magnetic field at the break from the conical and parabolic domains are equal to each other.

Consider the case of a constant Doppler factor in both accelerating and saturation regions.
Equating RHS of the Equation (23) by N24 and Equation (12) by R22, we get the relation
\begin{equation}
\frac{\eta\Gamma_\mathrm{max}}{\rho^2}=0.045\,\frac{\Omega_{r\nu1}^{p_1}}{\Omega_{r\nu2}^{p_2}}\left(r_\mathrm{br}\sin\theta\right)^{p_2-p_1}
\label{rel-1}
\end{equation}
that ensures the continuity of a magnetic field in a plasma proper frame in parabolic (Poynting dominated) and conical (plasma bulk kinetic energy dominated) domains. Here $\eta$ is a fraction of emitting plasma (see details in N24). Equation~(\ref{rel-1}) improves assessment of a magnetic field strength by fixing the value of $\eta\Gamma_\mathrm{max}/\rho^2$ term.  

Suppose now, that only the break in a core shift dependence on the frequency has been obtained, while the break in a jet form has been undetected due to the resolution reasons. In this case we presume that $k_1=k_\mathrm{r1}/2$ and $k_2=k_\mathrm{r2}$ (see details in R22 and N24). The position of a break is approximated by Equation~(\ref{r}). In this case we obtain
\begin{equation}
\frac{\eta\Gamma_\mathrm{max}}{\rho^2}=0.045\,\left(\frac{\Omega_{r\nu1}}{\Omega_{r\nu2}}\right)^{(2.5+\alpha)k_\mathrm{r2}}\nu_\mathrm{br}^{(2.5+\alpha)(1-k_\mathrm{r2}/k_\mathrm{r1})}.
\label{r1}
\end{equation}
Together with a break position assessment by Equations~(\ref{r}) and (\ref{d_proxy}) it again improves the magnetic field estimate.

For the case of a small enough viewing angle, when the Doppler factor in an accelerating domain depends on a Lorentz factor as $\delta\approx 2\Gamma$, we equate the magnetic field from Equations (25) by N24 and (\ref{LambdaSigma}) (in Equation~(25) there is a misprint, $\rho^2$ should be absent):
\begin{equation}
\frac{\eta\Gamma_\mathrm{max}d_\mathrm{br}^2}{R_\mathrm{L}^2}=0.045\delta^2\frac{\Omega_{r\nu1}^{p_1}}{\Omega_{r\nu2}^{p_2}}\left(r_\mathrm{br}\sin\theta\right)^{p_2-p_1},
\label{case2-1}
\end{equation}
where $\delta$ is a Doppler factor in the saturation domain. Hence, one can set $\delta=\Gamma_\mathrm{max}$.
Using the relation~(\ref{G}), we obtain exactly the Equation~(\ref{rel-1}), although the exponent $p_1$ in this case is different due to non-zero $k_{\delta}$ (see Equation~(\ref{p_power})).
Again, if the jet boundary shape has not been measured, the condition on the value of a parameter $\eta\Gamma_\mathrm{max}/\rho^2$ is given by Equation~(\ref{r1}).

Thus, matching the magnetic field at the jet shape geometry transition point restricts the model parameters with the universal Equations~(\ref{rel-1}) and (\ref{r1}) and allows for the more precise magnetic field estimate. Only the parabolic branch allows to extrapolate the magnetic field upstream the break. Simple extrapolation of $B_*$ calculated in the conical region into  parabolic part of a jet may provide incorrect result.

\section{Magnetic field in NGC~315}
\label{s:ngc315}

A jet from the source NGC~315 is unique, since both a break in a jet boundary shape and a break in frequency dependence of a core shift have been located by \citet{Boccardi_2021} and \citet{Ricci22}. Here we present the study of a magnetic field structure in NGC~315 jet using these results.  

Let us assume that we observe a jet in NGC~315 at the viewing angle $\theta\approx\Gamma^{-1}$ (the Doppler factor $\delta\approx\mathrm{const}$). We use the exponent $k_\mathrm{r1}=0.57\pm 0.17$ (R22) and the core shifts between ten pairs of frequencies 43.2, 22.3, 15.4, 12.1, 8.4~GHz \citep{Boccardi_2021} in quasi-parabolic domain. In quasi-conical region we use $k_\mathrm{r2}=0.93\pm 0.01$ and three pairs of frequencies 8.4, 5.0, 1.4~GHz. The corresponding core shift offsets are equal to $\Omega_{r\nu1}=13.2\pm 5.3 \;\mathrm{pc\, GHz^{1.75}}$ and $\Omega_{r\nu2}=2.58\pm 0.76\;\mathrm{pc\, GHz^{1.08}}$. The relation~(\ref{Jump}) holds for NGC~315 with a very good accuracy.

At first, let us check the approximations (\ref{r}) and (\ref{d_proxy}).
Setting the break frequency $\nu_\mathrm{br}=8.4$~GHz (R22), we obtain $r_\mathrm{br}=0.51\pm 0.20$~pc, which is close to the measured $0.58$~pc \citep{Boccardi_2021}. On the other hand, one can reverse the expression (\ref{r}) and find the break frequency, which for NGC~315 is equal to $7.8\pm 1.8$~GHz. Overall, the results for $\nu_\mathrm{br}$ by R22 and $r_\mathrm{br}$ by \citet{Boccardi_2021}  are consistent, while the estimated $r_\mathrm{br}$ by \citet{Park21} has a value much larger than the one, obtained by using $\nu_\mathrm{br}$. For the approximation (\ref{d_proxy}) we use $\varphi_\mathrm{app}=6^{\circ}.9$ by \citet{MOJAVE_XIV} and get $d_\mathrm{br}=0.038\pm 0.015$~pc. This estimate produces the value of a jet break width about two times smaller than the actual measured jet width at the break $0.072$~pc \citep{Boccardi_2021}. This difference may be connected with the error in estimation of apparent opening angle and with the uncertainty in the value of a viewing angle.

Using the Equation~(\ref{rel-1}) we obtain the condition ensuring the equality of magnetic fields, determined within the quasi-parabolic and quasi-conical domains:
\begin{equation}
\frac{\eta\Gamma_\mathrm{max}}{\rho^2}=5.25^{+29.75}_{-4.50}.
\label{cond_ngc315}
\end{equation}
The errors in a magnetic field estimates for different sets of observational frequencies provide 
the errors in the relation~(\ref{cond_ngc315}). This relation does not depend on the observed Doppler factor $\delta$, although the estimate of a magnetic field do depend on it as $\delta^{-0.5}$ (N24). Below we use $\delta=1$ for a large observational angle $\theta=38^{\circ}$ \citep{Boccardi_2021}.

To calculate the magnetic field in the quasi-parabolic accelerating part of a jet in NGC~315 we use the equations (23) from N24 and Equation (12) from R22. We observe that $\Omega_{r\nu1}$ is a few times larger than $\Omega_{r\nu2}$. Even with this difference, the estimates for a magnetic field upstream and downstream the break coincide for reasonable values of a fraction of emitting plasma $\eta$, maximum Lorentz factor $\Gamma_\mathrm{max}$ and coefficient $\rho$ given by (\ref{cond_ngc315}). Applying the condition~(\ref{cond_ngc315}), the magnetic field values in a plasma proper frame up and down the break are equal to $B_\mathrm{p*,\,br}=0.18\pm0.05$~G and $B_\mathrm{\varphi*,\,br}=0.18\pm0.04$~G correspondingly.

Let us extrapolate the magnetic field along a jet. In the domain of effective plasma acceleration one may adopt $B_*=B_\mathrm{p*}=B_\mathrm{p}$ \citep{Vlahakis04, Kom09}, since the plasma bulk velocity has a dominant poloidal velocity. Using the flux conservation, we obtain
for $r<0.58$~pc the following:
\begin{equation}
B_\mathrm{p*}(r)=B_\mathrm{p}(r)=B_\mathrm{p*,\,br}\left(\frac{r_\mathrm{br}}{r}\right)^{2k_1},\quad r<r_\mathrm{br}.
\label{Bp1}
\end{equation}

\begin{figure}
  \resizebox{\hsize}{!}{\includegraphics{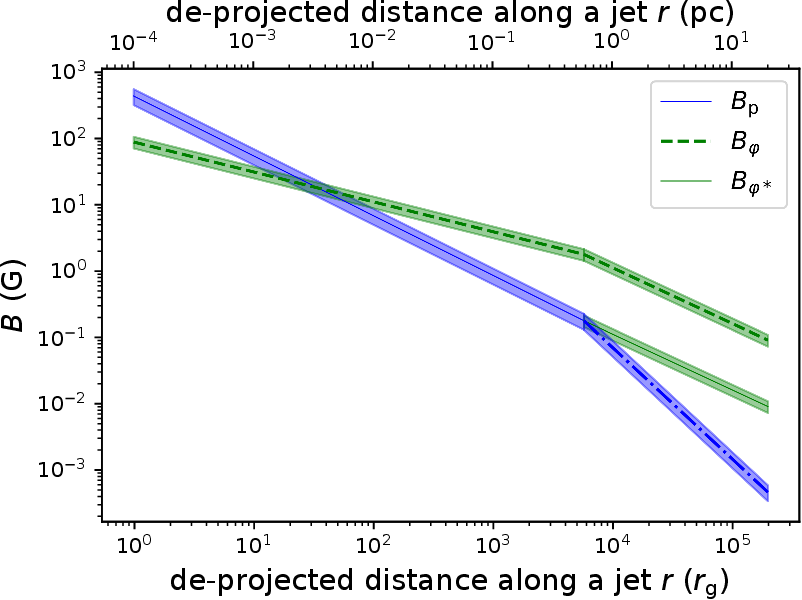}}
  \caption{Blue solid line --- a magnetic field poloidal component, calculated using the core shift measurements upstream the break. Shadowed region shows errors. Green solid line --- a magnetic field toroidal component in a plasma proper frame, calculated using the core shift measurements downstream the break. Blue dashed dotted line --- extrapolation of a poloidal field downstream the break. Green dashed line --- a magnetic field toroidal component recalculated into the core frame as $B_\varphi=B_{\varphi *}\Gamma_\mathrm{max}/2$. Upstream the break it was extrapolated using the dependence $B_\varphi\propto r^{-k_1}$. The toroidal magnetic field component in a plasma proper frame coincides with the poloidal component. Here we used the parameters $\Gamma_\mathrm{max}=20$, $\rho=0.2$, $\eta=0.011$. Distance along the jet is de-projected with the viewing angle $\theta=38^{\circ}$ (R22). A gravitational radius $r_{\mathrm{g}}$ corresponds to the black hole mass $M_\mathrm{BH}=2.08\times 10^{9}\,M_{\odot}$ \citep{Boizelle_21}.
  }
  \label{f:B1}
\end{figure}

On the other hand, measured core shift in quasi-conical domain provides the toroidal magnetic field $B_{\varphi *}$ in a plasma proper frame as a dominant component. We obtain
\begin{equation}
B_{\varphi*}(r)=B_\mathrm{\varphi*,\,br}\left(\frac{r_\mathrm{br}}{r}\right)^{k_2},\quad r>r_\mathrm{br}.
\label{Bhi1}
\end{equation}
Here $k_1=0.45$ is a power of a parabolic jet boundary shape $d\propto r^{k_1}$, and $k_2=0.84$ --- of a conical shape. The results are presented in Figure~\ref{f:B1}. The corresponding magnitudes of poloidal (blue) and toroidal (green) components in a plasma proper frame are shown in solid lines. As the powers of a magnetic field dependence on distance are equal to $2k_1=0.9$ and $k_2=0.84$, the difference in slopes are hardly noticeable. 

The approach of estimating magnetic field in both parabolic and conical domains allows us to assess its topology. First, the poloidal magnetic field can extrapolated downstream the break using the magnetic flux conservation:
\begin{equation}
B_\mathrm{p}=B_\mathrm{p*,\,br}(r_\mathrm{br}/r)^{2k_2},\quad r>r_\mathrm{br}
\end{equation}
(dashed dotted blue line in Figure~\ref{f:B1}). 
The proper frame toroidal field at the break $B_{\varphi *\,\mathrm{br}}$ relates to the field in a core frame $B_{\varphi,\,\mathrm{br}}$ as $B_{\varphi,\,\mathrm{br}}=B_{\varphi *,\,\mathrm{br}}\Gamma_\mathrm{max}/2$. Here we assume that a half of an initial Poynting flux converts into the the bulk plasma motion kinetic energy flux at the break. This means $\Gamma=\Gamma_\mathrm{max}/2$, and we adopt $\Gamma_\mathrm{max}=20$ by R22. From the break point at $r=0.58$~pc we extrapolate $B_\varphi$ upstream as 
\begin{equation}
B_\varphi=B_{\varphi *,\,\mathrm{br}}\frac{\Gamma_\mathrm{max}}{2}\left(\frac{r_\mathrm{br}}{r}\right)^{k_1},\quad r<r_\mathrm{br}
\label{Bphi_up}
\end{equation}
and downstream as 
\begin{equation}
B_\varphi=B_{\varphi *,\,\mathrm{br}}\frac{\Gamma_\mathrm{max}}{2}\left(\frac{r_\mathrm{br}}{r}\right)^{k_2},\quad r>r_\mathrm{br}
\label{Bphi_down}
\end{equation}
(dashed green line in Figure~\ref{f:B1}).
Extrapolating $B_{\varphi*}$ upstream the break is less trivial due to changing Lorentz factor of a bulk plasma motion:
\begin{equation}
B_{\varphi*}=B_\mathrm{\varphi*,\,br}\left(\frac{r_\mathrm{br}}{r}\right)^{2k_1},\quad r<r_\mathrm{br},
\end{equation}
where it coincides with the poloidal magnetic field upstream the break. Let us note that the radio core shift is fitted by R22 with respect to the the 43 and 86~GHz radio cores, while we use the distance $r$ from the jet base. However, this does not affect the result for a magnetic field extrapolation. The core shift offset, used to estimate $B_\mathrm{*,\,br}$, is calculated for pairs of frequencies. Extrapolation is performed from $r_\mathrm{br}$ obtained by the jet boundary shape modelling, which is not affected by the particular position of 43 or 86~GHz cores. 

\begin{figure}
  \resizebox{\hsize}{!}{\includegraphics{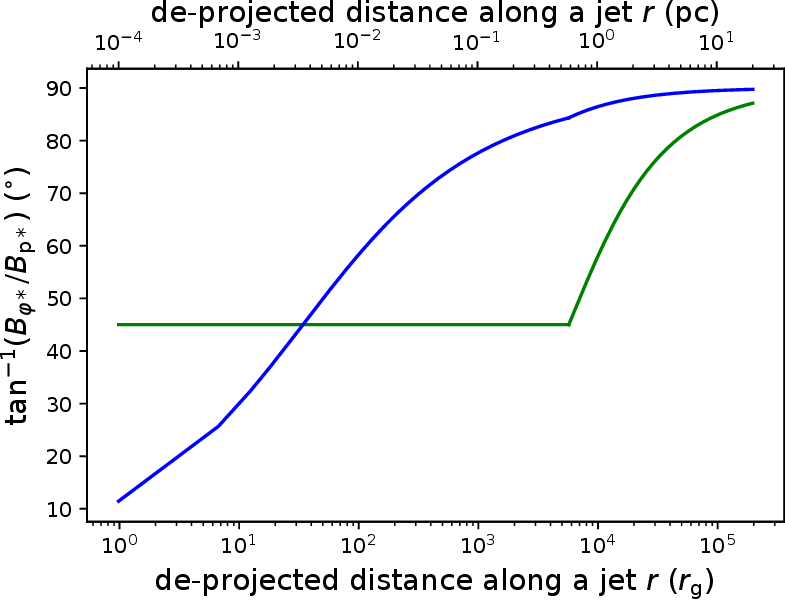}}
  \caption{Angle between the magnetic field and the direction along a jet in a plasma proper frame (green line) and in a core frame (blue line). Horizontal axis is the same as in Figure~\ref{f:B1}.}
  \label{f:alpha}
\end{figure}

In Figure~\ref{f:alpha} we present the angle between the direction of a magnetic field and the direction of a jet in a plasma proper frame (green line). The pitch angle of emitting particles is set with respect to this angle. The blue line in Figure~\ref{f:alpha} represents an angle between a magnetic field in a core frame and a jet direction. The magnetic field topology changes from the poloidal- to the toroidal-dominated along the jet.

Now let us assume that the viewing angle $\theta\ll\Gamma^{-1}$ (the Doppler factor $\delta\approx 2\Gamma$, $k_\mathrm{r1}=4k_1/3$). Using the relation (\ref{rel-1}) or (\ref{case2-1}) with different exponent $p_1$, than in the previous consideration due to $k_{\delta}=k$, we obtain
\begin{equation}
\frac{\eta\Gamma_\mathrm{max}}{\rho^2}=0.20^{+1.10}_{-0.17}.
\label{cond_2}
\end{equation} 
After ideal matching of magnetic field at the break, the picture of its distribution coincides exactly with the one, presented in Figure~\ref{f:B1}, but with amplitudes $\sqrt{\delta}=\sqrt{\Gamma_\mathrm{max}}$ times smaller. 

The value $\Gamma_\mathrm{max}=20$ is adopted from R22 (please, note, that in this paper $\Gamma_\mathrm{max}$ has a different meaning and is two times smaller) as the typical one for an expected parent population. The observed values of the bulk motion Lorentz factor in NGC~315 are smaller and equal to $\sim 3-4$ \citep{Ricci22, Park21}. However, for a large viewing angle, which is probably the case for NGC~315, we may not detect the fastest motion, which justifies using the greater $\Gamma_\mathrm{max}$. On the other hand, setting different maximum Lorentz factor would affect only the toroidal magnetic field in the core frame --- a green dashed line in Figure~\ref{f:B1}. Using $\Gamma_\mathrm{max}\approx 8$, which provides the maximum observed $\Gamma\approx 4$, makes an estimate of a toroidal magnetic field component in a jet $2.5=20/8$ times smaller, while the estimates for a magnetic field in a plasma proper frame would be the same due to Equation~(\ref{rel-1}).

The obtained magnetic field estimate is in a good agreement with the results in the recent work by \citet{Kino24}. Comparing the spectral index distribution with the model predictions \citep{Ro23}, they obtained the minimum magnetic field at $r=0.12$~pc ($1.6\times 10^3\;r_\mathrm{g}$ with $M_\mathrm{BH}=1.6\times 10^9\;M_{\odot}$) $B=0.8$~G, which is corroborated by the magnetic field estimate obtained above using the core shift break method. The blue solid curve in Figure~\ref{f:B1} coincides almost exactly with the lower boundary of the grey curve in Figure~5 by \citet{Kino24}. 
We should emphasise the difference of the methods in the paper by \citet{Kino24} and in this paper. \citet{Kino24} explore the region around the break point inside the parabolic domain, where the jet bulk motion acceleration undergoes a transition phase from ideal acceleration to saturation with a approximate power law $\Gamma\sim r^{0.3}$ . At the same time they estimate the toroidal magnetic field in a plasma proper frame \citep[see e.g.][]{Zamaninasabetal2014}, which is of the same order with the poloidal component in the parabolic domain. Both these assumptions lead to $B_*\propto r^{-0.88}$, very close to the assumed $B_*\propto r^{-0.9}$ for $k=0.45$ in this work. We, on the other hand, regard the domains of ideal linear acceleration and saturation, providing that core shift is measured separately between low (in conical) and high (parabolic domain) frequencies. These branches are determined for linear and saturated acceleration, but provide the result, which is close to the one obtained by \citet{Kino24}. Thus, the magnetic field behaviour, implied from studying the transition domain by \citet{Kino24} and two far asymptotic domains in this work demonstrates a very good agreement.

We see that having measured the core shift break (R22), we are able to match the estimates of the magnetic field $B_{*}$ calculated in both quasi-parabolic and quasi-conical jet domains. This is done for the reasonable, although not well restricted, values of a bulk plasma motion maximum Lorentz factor $\Gamma_\mathrm{max}$ and the fraction of emitting plasma $\eta$. Approximately the same results are obtained for different assumptions on a Doppler factor $\delta$. The developed here technique refines the method proposed by N24. 

\section{Discussion}
\label{s:Dis}

In general, we should expect the break in an apparent core position as a function of an observational frequency. This can take place even for the continuous behaviour of a jet emitting plasma number density amplitude $K_{e*}$ and magnetic field $B_*$ across a break. In a case of $\delta\approx 2\Gamma$ ($k_{r1}=4k_1/3$) the discontinuity in a core shift must be always present due to different powers in core shift in parabolic and conical regions. In a case of $\delta\approx\mathrm{const}$, the approximate equality of $k_\mathrm{r1}\approx k_\mathrm{r2}$, if it happens, may wipe out the effect. 

There can be two sources of a core shift break. The first one is a jump in $k_\mathrm{r}$, connected with the different regimes of a jet acceleration and / or jet different shape (see the upper and medium panels in Figure~\ref{f:cartoons}). The second one may occur due to the jump in parameters of a jet plasma and a magnetic field as a result of a plasma heating by its acceleration on shocks or due to a reconnection \citep[see e.g.,][]{Kirk1994,SironiSpitkovskyArons2013}. These two types can be distinguished by checking the relation~(\ref{Jump}), which must hold only for the first type of a core shift offset jump.
The first type of a jump in both $k_\mathrm{r}$ and $\Omega_{r\nu}$ due to the different regimes of a jet acceleration is observed in NGC~315 (R22), where the two branches intersect (see Figure~\ref{f:cartoons}, upper panel). In a case of $\Omega_{r\nu1}\le\Omega_{r\nu1}$ it can appear as two non-intersecting branches with different values of $k_\mathrm{r}$ (see Figure~\ref{f:cartoons}, the medium panel). The second type may appear as two non-intersecting branches as in the lower panel of the Figure~\ref{f:cartoons}. It is worth revisiting the multi-frequency core shift data, such as the one by \citep{Sokolovsky11}, to search for the possible core shift breaks. We will address this issue in the upcoming paper. 

The magnitudes of a poloidal and toroidal magnetic field in an AGN frame are of the same order only on the scales $d_*\sim R_\mathrm{L}$. This corresponds to the intersection of a blue solid and green dashed lines in Figure~\ref{f:B1}. Equating the amplitudes given by equations~(\ref{Bp1}) and (\ref{Bphi_up}) and using expression~(\ref{G}), we obtain the jet width $d_*=2R_\mathrm{L}/\rho$. Thus, for any choice of $\Gamma_\mathrm{max}$ the value of the light cylinder radius is self-consistent.

Measurement of both core shift and jet shape boundary break allows to refine the method of estimating a magnetic field in jets. Indeed, having the information on a core shift in both Poynting-dominated and plasma-dominated regions, we are able to limit the factor $\eta\Gamma_\mathrm{max}/\rho^2$ in expressions (23) and (25) by N24 and to improve the magnetic field estimates upstream the break. Potentially, both core shift and jet shape measurements allow to fit the break simultaneously, which will increase the precision of a break location.   
At the same time, the localisation of a break in a core shift allows to extrapolate the magnetic field onto the gravitational radius scales correctly, assuming the total magnetic flux conservation.

Conditions (\ref{cond_ngc315}) and (\ref{cond_2}) could be checked against the equipartition. The ratio of a magnetic to an emitting particles energy density changes along a jet for reasonable assumptions on $B_*(r)$ and $n_*(r)$ behaviour. In particular, for the dependencies~(\ref{defB}) and (\ref{defn}) with $2k_\mathrm{b}>k_\mathrm{n}$, the emitting plasma energy density grows along the flow, which may affect the brightness temperature in the cores \citep{Nok17A}. Let us introduce the ratio of a magnetic field to an emitting plasma energy density, which is conserved along a non-accelerating jet:
\begin{equation}
\Sigma_{\zeta}=\frac{B_{*\zeta}^2}{8\pi mc^2\Lambda K_{e*\zeta}}.
\label{Sigma}
\end{equation}
Here $\Lambda$ depends on the emitting plasma energy distribution. For $\alpha=0.5$ parameter $\Lambda=\ln(\gamma_\mathrm{max}/\gamma_\mathrm{min})$ and is usually set to be of the order of 10 \citep{L98}. With $\Sigma_{\zeta}$ and $\Lambda$ explicitly written, the expression (12) from R22 for a magnetic field in a conical non-accelerating jet is modified as follows:
\begin{equation}
B_{*\zeta}=0.0167\left[\frac{\Omega_{r\nu}^{p_2}}{r_{\zeta}^{p_2}}\left(\frac{1+z}{\delta}\right)^{(1.5+\alpha)}\frac{\Sigma_{\zeta}\Lambda}{d_{\zeta}\sin^{p_2-1}\theta}\right]^{1/(3.5+\alpha)}.
\label{LambdaSigma}
\end{equation}
Applying this latter equality to NGC~315, we obtain for $\delta\approx\mathrm{const}$
the relation $\Sigma_\mathrm{break}\eta\Gamma_\mathrm{max}/\rho^2=5.25^{+29.75}_{-4.50}$, which generalises the condition (\ref{cond_ngc315}). On the other hand, combining relation (\ref{Sigma}) with the equality (22)
by N24 we get
\begin{equation}
\frac{\Sigma_\mathrm{break}\eta\Gamma_\mathrm{max}}{\rho^2}=\frac{1}{\Lambda}.
\label{equip_cond}
\end{equation}
For the fiducial value of the parameter $\Lambda=10$, the generalised condition of continuity of a magnetic field in a plasma proper frame (\ref{cond_ngc315}) is fulfilled marginally. However, the generalised condition (\ref{cond_2}) is in a good agreement with the expected values for $\Lambda$. Assuming the equipartition in conical domain $\Sigma_\mathrm{break}=1$, we expect
the equality $\eta\Gamma_\mathrm{max}/\rho^2=0.1$.
This immediately leads to the small fraction of an emitting plasma with the upper limit $\eta\le 0.1/\Gamma_\mathrm{max}$. This result is suggested by \citet{Frolova23} basing on predicted intensity distribution. Here we must note again that the relation between an emitting plasma and magnetic field energy density (22) from N24 means, that a magnetic field dominates upstream the break. Its dominance falls as the flow propagates along a jet, and the classical equipartition holds starting from the jet shape break on. If there is no equipartition at the break, the ratio $\Sigma_\mathrm{break}$ still holds downstream. This can affect the measured brightness temperature of radio cores \citep{Pjanka17, Nok17A}.

It is non-trivial, that the condition of a magnetic field continuity (\ref{cond_2}) satisfies the independent condition (\ref{equip_cond}). 
The latter together with much better relation for a core shift exponents $k_\mathrm{r1}=4k_1/3$, may point to the non-constant Doppler factor regime in a parabolic domain (equation (25) by N24). As we noted earlier, this requires much smaller viewing angle $\theta\sim(\Gamma_\mathrm{max}/2)^{-1}$~radians. Such a value of a viewing angle places the break at the distance $3.2\;\mathrm{pc}=3.1\times 10^4\;r_\mathrm{g}$ for $M_\mathrm{BH}=2.08\times 10^9\;M_{\odot}$, which is still well within the estimated Bondi radius \citep{Boccardi_2021}.

The method of a core shift break may provide more sources with a detected break in a jet shape, but requires the multi-frequency observations.  
The typical frequencies of observations are 1, 5, 8, 12, 15, 22, 43, 86~GHz, with 230 and 345~GHz available with {\it EHT}, {\it ALMA} and planning {\it Millimetron} programs. In order to fit the core shift dependence with the two branches, two conditions must be fulfilled. 

First condition:
the break frequency must be $5-12$~GHz so that there are at least tree points to fit with every core shift branch. \citet{Kov-20, NKP20_r2} found that a geometry transition is located at the distance $10^5-10^6\;r_\mathrm{g}$. On the other hand, the position as close as few of $10^3\;r_\mathrm{g}$ may be present \citep{Boccardi_2021}. Below in our estimates we use the estimate $r_{\mathrm{br}}\approx 10^4-10^5\;r_\mathrm{g}$. We set the typical black hole mass as $10^9\;M_{\odot}$ and typical viewing angles $5^{\circ}$ for BL Lacs and $15^{\circ}$ for radio galaxies \citep[see details in][]{Kov-20}. This places the projected distance of a jet boundary shape break at the distance $0.04-1.3$~pc. Using Equation~(\ref{r}) with $\nu_\mathrm{br}=8$~GHz we obtain the lower limit for the value of a core shift offset $\Omega_{r\nu1}\gtrsim 0.9-9\;\mathrm{pc\; GHz^{1.5}}$ for BL Lacs, and $\Omega_{r\nu1}\gtrsim 3-30\;\mathrm{pc\; GHz^{1.5}}$ for radio galaxies, which makes the blazars preferable targets. We set here $k_\mathrm{r}=2/3$ for an accelerating jet with variable Doppler factor. 

The second condition is a sufficient resolution at high frequencies. Assuming the typical error in core shift measurements with ground-based VLBI $0.05$~mas \citep{Sokolovsky11, Pushkarev12}, we set the projected distance between cores at two different frequencies at $\sim 0.1$~pc for nearby sources. This leads to the lower limit for a core shift offset $\Omega_{r\nu1}\gtrsim 4-13\;\mathrm{pc\; GHz^{1.5}}$ for frequency pairs $8-15$ and $15-22$~GHz correspondingly. The higher the observational frequency, the greater value of a core shift offset must be in order to be resolved with the ground-based interferometry. We conclude that all the nearby sources with $\Omega_{r\nu}>1.5\;\mathrm{pc\; GHz}$, calculated for $k_\mathrm{r}=1$ between the frequencies $8.1$ and $15.4$~GHz \citep[see e.g.][]{Pushkarev12} (corresponding to $\Omega_{r\nu}>5\;\mathrm{pc\; GHz^{1.5}}$ for $k_\mathrm{r}=2/3$), may be targets for searching for the core shift break. Future {\it Millimetron} mission with the maximum resolution $\sim 10^{-4}-10^{-5}$~mas \citep{Millim21} will be able to easily track the jet core shift with $\Omega_{r\nu1}\sim 1-20$ up to the frequencies $\sim 300$~GHz, corresponding to the planned high frequency by {\it EHT}. As to lower values of core shift offset, with this resolution the core shift can be measured for $\Omega_{r\nu 1}\gtrsim 0.2$ a the frequency pair $86-230$~GHz in nearby sources, and for $\Omega_{r\nu 1}\gtrsim 1$ for the sources at red shift 1. Measuring core shifts with the ground-based VLBI at low frequencies and with {\it Millimetron} at high frequencies may potentially help to detect more core shift breaks in nearby sources, and collect the data for core shift dependence on the observational frequency close to a jet base even for distant sources. 

It would be most beneficial to search for a core shift break in sources with already detected jet shape geometry break. However, analysis shows that the expected $\nu_\mathrm{br}$ (see Equation~(\ref{r})) has a value $\lesssim 1$~GHz for all the sources in a sample from N24, except for NGC~315 investigated here and the sources 1637$+$826 and 1807$+$698 with the break frequency expected at 3 and 4~GHz correspondingly. We do not apply the method of a core shift break to the jet in M87, since the radio cores at the observational frequencies from $230$ to $5$~GHz lie within the parabolic domain. However, applying the Equation~(23) by N24, applicable for an accelerating flow, we obtained the results in a very good agreement with the results by \citet{EHT_B}. Again, our method provides the result in excellent agreement with \citet{Ro23} but a few times smaller. Thus, for the sources with a detected jet shape break and expected $\nu_\mathrm{br}<1$, it is still important to conduct multi-frequency observations to study the effects of both flow acceleration and viewing angle on a core shift.

\section{Conclusions}
\label{s:Conc}

The work is dedicated to the presence of a change in core shift exponent $k_\mathrm{r}$ and, consequently, to the jump in a core shift offset $\Omega_{r\nu}$ at the jet geometry transition region $r_\mathrm{br}$. As the jet changes its shape from parabolic to conical and / or plasma bulk motion acceleration --- from effective to saturation regime, the opacity behaves differently in two asymptotic regions. Although the dependence of a radio core position on an observational frequency is continuous, it has two asymptotic branches, corresponding to two different $k_\mathrm{r}$ and $\Omega_{r\nu}$.

1. We show that, in general, the core shift offset $\Omega_{r\nu}$ must be discontinuous in the region, where the jet boundary changes its shape. In particular, it is present in the region where the plasma bulk motion pattern transits from the effective acceleration, given by the first line in Equation (\ref{G}), to the extremely slow acceleration downstream. It is also present if the jet changes its shape within one acceleration regime, or if plasma bulk motion Lorentz factor saturates without a change in a jet boundary geometry.
In all these cases the relation~(\ref{Jump}) must hold for a jump in measured core shift offset.

2. We discuss the different possible forms of a jump in core shift vs. observational frequency relations presented by the cartoons in Figure~\ref{f:cartoons}. We speculate, that the balance between a magnetic field and an emitting plasma energy density may change abruptly as a consequence of a plasma acceleration at a shock or as a result of a reconnection event. This will also lead to a jump in $\Omega_{r\nu}$, given by Equation~(\ref{Flare}). If such an event takes place at the distances of a radio core at the frequencies $15-22$~GHz, as expected in blazars possibly demonstrating a high-energy neutrino production \citep{Plavin-20}, it could be potentially detected by a dedicated search. Such events in general would produce a core shift offset jump that can be differentiated from the discussed thoroughly in this paper. If the break in a core shift appears due to plasma heating or a flare propagation, than the core shift exponent must be the same, and the value of a jump in general would not obey the relation~(\ref{Jump}).

3. Measuring both branches of a core shift offset allows limiting the parameter $\eta\Gamma_\mathrm{max}/\rho^2$ by using Equations (\ref{r1}) or (\ref{case2-1}). Thus, the detection of a core shift break refines a magnetic field estimate in the parabolic part of a jet. The latter may be extrapolated upstream the break and, possibly, onto the scales of a gravitational radius.

4. We have demonstrated on the example of NGC~315 how using both branches of a core shift may improve the precision of a magnetic field estimate. The break in a core shift together with a jet shape break can also refine estimation of the parameters of a jet geometry transition region.

5. Applying different cases of a relation between the observation angle and bulk motion Lorentz factor, we find that the case $\delta\approx 2\Gamma$ explains much better the rich observational data for NGC~315. However, it means that the observational angle of a jet must be of the order of $6^{\circ}$. Even for such small angle, the de-projected distance of a jet shape break lies well within the estimated Bondi sphere.

6. We propose the core shift break search as a possible new instrument for the location of a jet shape break in the larger sample of sources than available at the moment. The core shift measurements at high frequencies with the resolution provided by the space-ground interferometry will allow studying the jet acceleration regions even for high red shift sources.

\section*{Acknowledgements}

We thank the anonymous referee for thoughtful comments and suggestions that helped to
improve the paper.
The work was supported by the Foundation for the Advancement of Theoretical
Physics and Mathematics ``BASIS''. This research made use of the data from the MOJAVE database\footnote{\url{http://www.physics.purdue.edu/MOJAVE/}} which is maintained by the MOJAVE team \citep{MOJAVE_XV}.
This research made use of NASA's Astrophysics Data System.

\section*{Data availability}

There is no new data associated with the results presented in the paper. All the previously published data has the proper references.

\bibliographystyle{mnras}
\bibliography{nee1}

\appendix

\bsp    
\label{lastpage}

\end{document}